\documentclass[12pt]{iopart}

\usepackage{graphicx}

\newcommand{\cfigl}[3]{\begin{figure}[!hbtp]\centering% !ht 
 \includegraphics[width=6cm]{#2}\caption{\small{{#3}}}\label{#1}\end{figure}}

\usepackage{iopams}

\begin{document}
%\draft
\title{The center of mass and center of charge of the electron}
\author{M.\ Rivas}
\address{Teoretical Physics Department, University of the Basque Country,
Apdo. 644, 48080 Bilbao (Spain)}
\ead{martin.rivas@ehu.es}

%\date{\today}

%\maketitle

\begin{abstract}
If the assumption that the center of mass(CM) and the center of charge(CC) of the electron  
are two different points was stated 100 years ago, our conceptual ideas about elementary particles 
would be different. This assumption is only compatible with a relativistic description. It suggests, 
from the classical point of view, that the angular momentum of the electron has to have a unique value.
In the free motion, the CC follows a helix at the speed of light. 
The spin with respect to CC and to CM satisfy two different
dynamical equations which shows that Dirac spin operator in the quantum case satisfies the same dynamical
equation as the classical spin with respect to the CC. This means, among other things, that the addition of the three
Dirac's spin operators of the three quarks can never give rise to the spin of the proton, so that the proton 
spin crisis could be related to this incompleteness in the addition of the quark's angular momenta. Some
other effects related to spinning particles like the concept of gyromagnetic ratio, 
a classical description of tunneling and the formation
of bound pairs of electrons, are analized.
\end{abstract}

%\vspace{2cm}

%%%%%%%%%%%%%%%%%%%%%%%%%%%
\section{Introduction}
\label{sec:intro}
%%%%%%%%%%%%%%%%%%%%%%%%%%%%%%%%%%%%

If real particles are exactly mathematical points then all their mechanical and electromagnetic properties 
will be defined at that point. If they are not points, mechanics defines for any material system 
a geometrical point ${\bi q}$, called the center of mass, such that the linear momentum of 
the system takes the form ${\bi p}=m{\bi v}$, 
where ${\bi v}=d{\bi q}/dt$ in a nonrelativistic framework, or ${\bi p}=\gamma(v)m{\bi v}$, 
in a relativistic one, where $\gamma(v)=(1-v^2/c^2)^{-1/2}$. 
From the electromagnetic point of view, its electromagnetic structure can be reduced
to a single point ${\bi r}$ where we locate the total charge of the particle and the electric
and magnetic multipoles. This point is in general arbitrary, but if the charge distribution has a complete
spherical symmetry around a point, the electric structure around that point is reduced to the value of the 
total charge and no other electric multipoles. 
If the current densities are also spherically distributed in a radial
direction around that point, the magnetic multipoles also vanish. Otherwise, nonradial currents will 
give rise to different magnetic multipoles.

We do not know what is the exact electromagnetic structure of the electron, how the charge and its 
possible internal currents are distributed. The usual coupling of quantum electrodynamics $j^\mu A_\mu$,
between the particle current $j^\mu$ and the external potentials $A_\mu$, is obtained by 
the local gauge invariance hypothesis.
We get that no multipole interactions with the external fields  appear in this coupling. 
For a strict point particle, that coupling is
reduced in the classical description to $-e\phi+e{\bi u}\cdot{\bi A}$, where ${\bi u}$ is the velocity of
the point, and $\phi$ and ${\bi A}$ the external scalar and vector potentials, respectively, defined at
that point. Conversely, if we describe the classical interaction of the electron with an external 
field in this form, we are explicitely assuming a strict spherical symmetry for the charge 
and current distribution.

In the next section we recreate a plausible academic situation about the classical description of a massive
charged particle under the hypothesis of two different centers. Before quantum mechanics this hypothesis
would be certainly rejected. But if this classical analysis formed part of the usual academic 
exercises and tripos at the universities, and its conclusions, although considered unphysical, 
were known to physicists, the early steps of quantum
mechanics would have been completely different and the quantization of the angular momentum of matter
would be assumed as a natural consequence of this assumption.

In section \ref{history} we review in a simplified way, several historical papers by very well known physicists,
and compare the different proposals with this assumption of two different centers. Section \ref{kinematic} 
is devoted to a summary of
a general formalism for describing elementary spinning particles, classical and quantum mechanical, proposed
by the author, in which the existence of two different points is not postulated, but it is a consequence of the
formalism. What is postulated is a physical definition of elementary particle. In this section we also describe
some physical effects that can have a clear classical explanation in this formalism, and where the separation
between the center of mass and the centre of charge plays an important role in its physical interpretation.

%%%%%%%%%%%%%%%%%%%%%%%%%%%
\section{A plausible academic story. Marianne's thesis}
\label{sec:story}
%%%%%%%%%%%%%%%%%%%%%%%%%%%

In the decade of 1910 a university professor proposes to the students the analysis of 
a masive charged particle interacting with an external electromagnetic field, 
but under the framework of the recently new special relativity theory. 
The analysis has to be done in a Lagrangian framework.
The postulated Lagrangian of the particle under some external electromagnetic field will be
\[
L=L_0+L_I,\quad L_I=-e\phi(t,{\bi r})+e{\bi u}\cdot{\bi A}(t,{\bi r}), \quad {\bi u}=\frac{d{\bi r}}{dt},
\]
where $e$ is the charge of the particle, $\phi(t,{\bi r})$ and ${\bi A}(t,{\bi r})$ are the external scalar 
and vector potentials, respectively, defined at the point ${\bi r}$ and $L_0$ the new relativistic free 
Lagrangian of a particle of mass $m$. 

One of the students, called Marianne, takes the subject and makes the following {\it ansatz}:

$-$ Sir, the Lagrangian $L_0$ describes the mechanical properties of the particle and $L_I$ its interaction.
This last part $L_I$ suggests that the particle is an object with a complete spherical symmetry as far as 
the charge distribution and the possible internal currents are concerned. 
Its complete electromagnetic structure is reduced to a charge $e$ located at a single point ${\bi r}$, where
the external fields are defined, and no other electric or magnetic multipoles, since there are no multipole
couplings in $L_I$. 
For another arbitrary point ${\bi k}$, the electromagnetic structure of the particle 
will be reduced to the same charge $e$ located at ${\bi k}$
and also an electric dipole ${\bi d}=e({\bi r}-{\bi k})$ and a 
magnetic dipole $\bmu=e({\bi r}-{\bi k})\times{\bi w}/2$
where ${\bi w}$ is the relative velocity of the point ${\bi r}$ with respect to the point ${\bi k}$, 
and no further multipoles. Therefore we can call the point ${\bi r}$, where the electromagnetic structure takes
its simplest form, the center of charge of the particle (CC).

From $L_0$, the different mechanical properties are defined. For instance the mechanical energy $H$ and the 
mechanical linear momentum ${\bi p}$, which, according to the special relativity 
are expressed as $H=\gamma(v)mc^2$ and ${\bi p}=\gamma(v)m{\bi v}$, respectively. The vector 
${\bi v}=d{\bi q}/dt$ represents the velocity of the center of mass ${\bi q}$ and 
$\gamma(v)=(1-v^2/c^2)^{-1/2}$. 
If the particle is exactly a mathematical point then ${\bi q}={\bi r}$, 
but if it is not exactly a point they could be 
different. The interacting part assumes a spherical symmetry of its electromagnetic structure, 
but the mechanical part says nothing about the mass distribution. 
Only the existence of a point ${\bi q}$ which represents 
the location of the CM. 
My proposal is to try to solve the problem by considering that both points are different.
 
$-$ OK lady Marianne, your proposal is more general. The assumption that real particles are 
exactly mathematical points is certainly an approximation. The dynamics will be more complicated 
because you have to describe the evolution of two points. Nevertheless, go ahead and we shall see what
kind of conclusions you reach, was the tutor's answer.

The next two subsections are a summary of Marianne's thesis.

%%%%%%%%%%%%%%%%%%%%%%%%%%%%%%%%%%%
\subsection{The free motion of the center of mass and center of charge}
\label{sec:spin}
%%%%%%%%%%%%%%%%%%%%%%%%%%%%%%%%%%%

From the $L_0$ part we get the free dynamical equation $d{\bi p}/dt=0$, and from the $L_I$ part 
the external Lorentz force, such that the complete dynamical equation will be
\[
\frac{d{\bi p}}{dt}=e({\bi E}(t,{\bi r})+{\bi u}\times{\bi B}(t,{\bi r})),\quad {\bi E}=-\nabla\phi-\frac{\partial{\bi A}}{\partial t},\quad {\bi B}=\nabla\times{\bi A}.
\]
The fields are being evaluated at the point ${\bi r}$. 
The left hand side is the time derivative of ${\bi p}=\gamma(v)m{\bi v}$, 
where ${\bi v}=d{\bi q}/dt$ is the velocity
of the CM. To obtain the trajectory of the CM we need to know the trajectory of the CC to evaluate the external force.
But the CC position ${\bi r}$ will be in the neighborhood of ${\bi q}$, close to it, 
so we must make some {\it ansatz} about their relationship. 
Let us see first how they are related in the free case.

If the particle is free ${\bi p}$ is conserved so that the CM ${\bi q}$
moves along a straight line with a constant velocity ${\bi v}$. But, what about 
the trajectory of the point ${\bi r}$?
 
From the geometrical point of view in three-dimensional space, the trajectory of a point which follows
a continuous and differentiable path, can be described as the evolution of its Frenet-Serret triad. 
This triad is displaced an arc length $ds=u dt$ in time $dt$ along the unit tangent vector ${\bi t}$, 
and rotates an angle $\kappa ds$ around the binormal ${\bi b}$, and also an angle $\tau ds$ around 
the tangent ${\bi t}$ in the same time, where $\kappa$ and $\tau$
are the instantaneous curvature and torsion of the trajectory, respectively.
But if the motion is free it means that we cannot obtain a different dynamical behaviour of the CC 
at two different times. Otherwise, a different dynamical behaviour will mean that something
different is happening at different times. 
The above infinitesimal displacement and angles must be independent of the 
time so that the CC follows a trajectory of constant curvature and torsion at a velocity of 
constant absolute value $u=ds/dt$.
The CC must follow a helix at a constant speed and this description must be valid for any inertial observer.
The CM seems to be the central point of the helix, i.e., the CM describes the axis of the helix,
such that the projection of the velocity ${\bi u}$ of the CC along the axis will be the CM velocity ${\bi v}$.

If the particle has two different points ${\bi q}$ and ${\bi r}$ the above description is incompatible with
a nonrelativistic treatment because if the helix is run at a constant speed for some particular
inertial observer then it is not described at a constant speed for a moving one,
if the composition of velocities is the usual vector addition. 
Because this motion is accelerated we can never find any inertial observer at rest with respect to the
CC. If at time $t$ one inertial observer is at rest and thus measures $u=0$, at time $t+dt$ the CC is moving 
with $u\neq0$ with respect to this observer
and this is contradictory with the condition that the motion is at a constant speed in that frame. 
The CC velocity has to be
an unreachable velocity of constant value for any inertial observer. Twenty years ago we would say that this is
impossible, but according to the special theory of relativity the speed of light is a good candidate for this unreachable
velocity limit.

In the special relativity, if the point ${\bi r}$ moves at the speed $c$ for an inertial observer,
then it moves with the same speed $c$ for all of them.
Only a relativistic description is compatible with the
assumption of two different points CC and CM, 
but necessarily the CC has to be moving at the speed of light $c$.
The free motion of a particle with two centers, implies that the CC follows a helix at 
the speed of light. In this case the CM velocity ${\bi v}$ will be the projection of the velocity
${\bi u}$ along the axis of the helix, and therefore $v<c$.

In Frenet-Serret analysis, the most general differential equation satisfied by a point in 
three-dimensional space, is the fourth order equation:
\[\fl
\frac{d^4{\bi r}}{ds^4}-\left(\frac{2\dot{\kappa}}{\kappa}+\frac{\dot{\tau}}{\tau}\right)\frac{d^3{\bi r}}{ds^3}
+\left(\kappa^2+\tau^2+\frac{\dot{\kappa}\dot{\tau}}{\kappa\tau}+\frac{2\dot{\kappa}^2
-\kappa\ddot{\kappa}}{\kappa^2}\right)\frac{d^2{\bi r}}{ds^2}
+\kappa^2\left(\frac{\dot{\kappa}}{\kappa}-\frac{\dot{\tau}}{\tau}\right)\frac{d{\bi r}}{ds}=0,
\]
where the over dot means $d/ds$, such that if the curvature and torsion are constants, it is reduced to
the linear differential equation with constant coefficients
\[
\frac{d^4{\bi r}}{ds^4}+(\kappa^2+\tau^2)\frac{d^2{\bi r}}{ds^2}=0.
\]
Since $ds=c dt$, in terms of time evolution it becomes
\[
\frac{d^4{\bi r}}{dt^4}+c^2(\kappa^2+\tau^2)\frac{d^2{\bi r}}{dt^2}=0,\quad \frac{d^4{\bi r}}{dt^4}+\omega^2\frac{d^2{\bi r}}{dt^2}=0,
\] 
where the constant positive coefficient $\omega$ has dimensions of time$^{-1}$. This coefficient, which must be
constant for all inertial observers, does not take the same value in every frame, because depends on the values of the
curvature and torsion of the curve in the corresponding frame. 
This equation can be recast in the following form in every frame 
\[
\frac{d^2}{dt^2}\left(\frac{1}{\omega^2}\frac{d^2{\bi r}}{dt^2}+{\bi r}\right)=0,
\]
so that the point ${\bi q}={\bi r}+\ddot{\bi r}/\omega^2$ satisfies $\ddot{\bi q}=0$, where now the overdot means $d/dt$,
and could be interpreted as the center 
of the helix, or the CM of the particle. In the center of mass frame ${\bi q}=0$ and thus $\omega$ is the angular
velocity of the point ${\bi r}$ around the CM, which describes a central motion at a constant velocity, i.e., a 
circle of radius $R=c/\omega$. If we use this definition to express the CM position in terms of ${\bi r}$ in the
Lagrangian $L_0$, we have a mechanical system described by a single point ${\bi r}$, the location of its CC, 
but the Lagrangian has to depend 
on the acceleration of the point ${\bi r}$ to obtain fourth order differential equations. It seems that the CC 
rotates with a unique angular velocity around the CM, but this angular velocity has to be a function
of the CM velocity.
 
{\bf Marianne's conclusions}:

This description is incompatible with a nonrelativistic framework, but not in the new special
relativity context. The hypothesis of two separate points requires the existence of velocities
of physical points, constant and unreachable for all inertial observers.
Although the CM does not move at the speed of light, the CC does. It is accelerated and therefore
the particle has to radiate. From the CM point of view the particle will behave though it has a 
magnetic moment because there exists a relative velocity between ${\bi r}$ and ${\bi q}$ 
and also an oscillating electric dipole
and this seems contradictory with the assumption that the particle has no dipole structure. 
The electromagnetic field created by this particle, from its center of mass at rest, will be a static Coulomb field,
the static magnetic field of a magnetic dipole at the origin and a non static electromagnetic field created
by a rotating electric dipole in a plane orthogonal to the magnetic moment. 

The particle
rotates and therefore it will have angular momentum. If the description is done in the CM frame, the motion
of the CC will be a circle of constant radius $R$, and this implies a constant and unique angular velocity
$\omega=c/R$ for all identical particles at rest, like the electrons, 
and also a constant and unique angular momentum,
which will be conserved in the free case. Otherwise, if the particle radiates, the radius will decrease and if the
CC velocity remains constant the angular momentum of the particle must also decrease. 

We know that for any rigid body any external torque modifies its angular momentum. In the case of this 
free particle at rest, no torque would modify the absolute value of the angular momentum. Only perhaps its orientation.
If there is no radiation, this unique internal motion implies that we 
can never modify the angular momentum of the particle. All identical particles, if its internal structure has not been
modified, when they are free must have the same 
constant angular momentum, irrespective of the interaction undergone. 
Otherwise, external torques would modify its internal structure.

%%%%%%%%%%%%%%%%%%%%%%%%%%%%%%%%%%%
\subsection{The angular momentum of the particle}
\label{sec:spin}
%%%%%%%%%%%%%%%%%%%%%%%%%%%%%%%%%%%

The angular momentum of any mechanical system is a magnitude which is defined with respect to some
prescribed point.
If the particle has two different characteristic points, the center of charge ${\bi r}$ and
the center of mass ${\bi q}$, the angular momentum of the particle can be defined with respect to both
points. Let us call ${\bi S}$ the angular momentum with respect to the CC and ${\bi S}_{CM}$ 
the angular momentum with respect to the CM. Even more, let us assume that ${\bi k}$ is another
point of the particle, different from ${\bi q}$ and ${\bi r}$ and let us call ${\bi S}_k$ 
the angular momentum with respect to this point.

Let ${\bi J}$ be the total angular momentum of the particle with respect to the origin of observer's frame.
It can be written in the following alternative forms, either
\[
{\bi J}={\bi r}\times{\bi p}+{\bi S},\quad {\rm or}\quad {\bi J}={\bi q}\times{\bi p}+{\bi S}_{CM},
\quad {\rm or}\quad {\bi J}={\bi k}\times{\bi p}+{\bi S}_{k},
\]
where ${\bi p}$ is the linear momentum of the particle in this frame. 
If the linear momentum for that observer vanishes then the angular momentum with respect to any point
in that frame takes the same value, so that ${\bi S}={\bi S}_{CM}={\bi S}_k$, for the center of mass observer.

If the particle is under some external
force applied at ${\bi r}$, ${\bi J}$ and ${\bi p}$ are not conserved but satisfy 
\[
\frac{d{\bi p}}{dt}={\bi F},\quad \frac{d{\bi J}}{dt}={\bi r}\times{\bi F}.
\]
Taking the time derivatives of the three expressions of the total angular momentum we arrive to
\[
\frac{d{\bi S}}{dt}={\bi p}\times\frac{d{\bi r}}{dt},\quad \frac{d{\bi S}_{CM}}{dt}=({\bi r}-{\bi q})\times{\bi F},\quad
\frac{d{\bi S}_k}{dt}={\bi p}\times\frac{d{\bi k}}{dt}+({\bi r}-{\bi k})\times{\bi F},
\]
because the linear momentum is pointing along the direction of the velocity of point ${\bi q}$ 
and not in general along the direction of the velocities of the other two points ${\bi r}$ and ${\bi k}$.
These three angular momenta satisfy three different dynamical equations.
In the free case ${\bi S}_{CM}$ is conserved while ${\bi S}$ and ${\bi S}_k$ evolve in a direction orthogonal
to ${\bi p}$, so that only their projections on ${\bi p}$, ${\bi S}\cdot{\bi p}$ and ${\bi S}_k\cdot{\bi p}$,
respectively, are conserved. 

The dynamical equation of the angular momentum ${\bi S}$ is independent of whether
the particle is free or not. Its evolution is always orthogonal to ${\bi p}$. 
This means that if we compute the angular momentum of the particle 
with respect to some point, the dynamical behaviour of this angular momentum will give us 
information about whether this point is the center of charge, the center of mass or a 
point different from them.

For the center of mass observer ${\bi p}=0$, and only the ${\bi S}$ is conserved.
If the particle is not free but both points are the same ${\bi r}={\bi q}$, then ${\bi S}_{CM}={\bi S}$ are 
conserved.
But, conversely, if the angular momentum with respect to the point where the external 
force is applied is not conserved it means that this point is a different 
point than the center of mass.

{\bf Marianne's conclusions}:

The CC angular momentum ${\bi S}$ satisfies an awful dynamical equation. It is not conserved for a free
particle. It seems to preccess
around the direction of the linear momentum and it is independent
of the external force. It is not the usual torque equation like the one satisfied by the total
${\bi J}$ or the ${\bi S}_{CM}$. The existence of a rich variety of forms of matter seems to be
contradictory with this restricted and unique internal motion associated to this particle. 
The analysis of a particle with two centers produces a unique
description of a motion of the CC at the speed of light. 
Why the CC of matter has to be moving at an unreachable velocity for inertial observers 
at such a small scale?
Are there any physical evidence of this fast motion? 
Are there any experimental evidence that all identical particles like electrons, have a unique and the
same angular momentum at rest? It seems that the assumption
of two separate centers is unphysical.\\

Most probably, this work done in 1910 by a lady, would be considered as an academic exercise without any 
physical interest because it describes unusual and unmeasured properties of matter at that time. 
This work would obtain average marks, not recommended for publication and its single
copy packed in a folder and kept inside a drawer in professor's office, if not burned. %The thesis I mean, not Marianne.

We have no evidence that this academic story would ever happened. 
But her analysis in terms of elementary differential geometry could certainly be done in that time.
It shows, from a classical point of view, that the angular momentum of a particle with two 
centers is not quantized, but rather that it takes a unique value. That magnitude, with dimensions of angular momentum, will become a 
universal physical constant.

The coming years were full of new physical phenomena and theoretical proposals, 
with very close resemblances to Marianne's conclusions.

%%%%%%%%%%%%%%%%%%%%%%%%%%%%%%%%%%%%
\section{Some findings and proposals in the next decade}
\label{history}
%%%%%%%%%%%%%%%%%%%%%%%%%%%%%%%%%%%%

According to classical electrodynamics any bound system of charged particles is unstable because 
if every particle is accelerated it must necessarily radiate.
Classical electrodynamics is not a complete theory. It does not justify the stability of matter.
But real matter seems to be stable. All matter must radiate until it reaches a final
stable stationary bound state. Classical electrodynamics needs some extra assumptions to describe reality.

%%%%%%%%%%%%%%%%%%%%%%%
\subsection{N. Bohr}
This is the idea that Niels Bohr proposed in 1913 \cite{Bohr}, to justify the stable 
structure of normal matter and in particular the simplest bound state
of two particles, the hydrogen atom. Although electrons in an atom are accelerated, there must exist some states
which under certain conditions, the atom will be in a stationary state. The state of lowest energy
will be the ground state of the system. The additional condition was that the orbital angular momentum of the 
electron in a stationary state, has to be necessarily a multiple of $\hbar$. 
The orbital angular momentum of the electron in an atom has to be quantized.
Without this assumption, the classical electromagnetic theory leads to non stable matter.

%%%%%%%%%%%%%%%%%%%%%%%
\subsection{L. de Broglie}
In 1924 Louis de Broglie defends his thesis, being published in 1925 \cite{Broglie}. 
He is known for the statement of the wave nature of the electron or the wave-particle duality of matter. 
Nevertheless, in de Broglie's thesis
there is not a single mention of this duality. 
The only extra assumption is: "{\it \ldots nous admettons dans le pr\'esent travail
l'existence d'un ph\'enom\`ene periodique d'une nature encore \`a pr\'eciser qui serait 
li\'e \`a tout morceau isol\'e d'\'energie et qui d\'ependrait de sa masse propre 
par l'\'equation de Planck-Einstein.}", i.e.,
"{\it \ldots we assume in the present work,
the existence of a periodic phenomenon
of an unknown nature, associated to every portion of isolated energy, and related to 
its proper mass by Planck-Einstein's equation.}" (author's translation). 
The frequency $\nu$ is given by $h\nu=mc^2$, so that a particle of mass $m$ at rest should have an internal
frequency $\nu_0=mc^2/h$ and when the particle moves at a velocity $v$ this frequency will be $\nu=mc^2/h\gamma(v)$, 
according to the time dilation of special relativity. 

%%%%%%%%%%%%%%%%%%%%%%%
\subsection{A.H. Compton, G. Uhlenbeck and S. Goudsmith}
In 1926 Uhlenbeck and Goudsmith publish a letter in {\it Nature} \cite{Uhlenbeck} where they mention
that previously in 1921, Compton \cite{Compton} suggested the idea of the spinning electron
as the origin of the natural unit of magnetism. In Compton's work he mentions: 
"{\it Let us then assume with Bohr
that if each electron has some definite angular momentum such as $h/2\pi$, no radiation occurs}", 
where the Compton
hypothesis of an angular momentum of the electron of integer value $\hbar$ is simply stated to 
prevent radiation of an hypothetical rotation of the electron.
Uhlenbeck and Goudsmith apply this idea to justify the fine structure and Zeeman effect of hydrogen-like atoms, but the 
value of the angular momentum of the stationary states S,P,D,\ldots is 1/2, 3/2, 5/2,\ldots, respectively.
With the hypothesis of a spinning electron of angular momentum $\hbar/2$, each level is separated into two, thus producing
a total angular momentum $J$ of integer value 1,2,3,\ldots. The electron has angular momentum. It can be 
interpreted that with the electron at rest it has angular momentum with respect to its center of mass. 
Remark that the total angular momentum is quantized in a wrong way.

The de Broglie and Uhlenbeck and Goudsmith assumptions are completely equivalent 
to assume that elementary matter rotates with a unique angular velocity, which
is related to its rest mass by Planck-Einstein's equation. 
If all matter that sourrounds us moves and rotates, why not to consider that
elementary matter also rotates? If Bohr has assumed a quantized orbital angular momentum 
for an accelerated particle under some external force, why the assumption
of a quantized angular momentum of a rotating free electron with respect to its CM 
is so strange? Bohr, at the end of Uhlenbeck and Goudsmith letter, comments: 
{\it "Having had the opportunity of reading this interesting letter
by Mr. Goudsmith and Mr. Uhlenbeck,\ldots, the introduction of the spinning electron which, in 
spite of the incompleteness
of the conclusions that can be derived from models, promises to be a very welcome supplement to our
ideas of atomic structure."}

%%%%%%%%%%%%%%%%%%%%%%%
\subsection{P.A.M. Dirac}
In his original papers in 1928 Dirac \cite{Dirac1, Dirac2} analyzes a system whose total energy and linear momentum
are, $H=H_m+e\phi$ and ${\bi p}={\bi p}_m+e{\bi A}$, where ${\phi}$ and ${\bi A}$ are the external potentials 
and $H_m$ and ${\bi p}_m$ the mechanical energy and linear momentum, respectively. 
Because the mechanical energy and linear momentum of a particle of mass $m$ satisfy
$(H_m/c)^2-{\bi p}_m^2=m^2c^2$, the total Hamiltonian and total linear momentum satisfy
\[
\frac{1}{c^2}\left(H-e\phi\right)^2-\left({\bi p}-e{\bi A}\right)^2=m^2c^2.
\]
After some algebraic manipulations, Dirac transforms this expression to get a linear function
of $H$, and therefore also of ${\bi p}$, and the Hamiltonian is finally described as
\[
H=({\bi p}-e{\bi A}(t,{\bi r}))\cdot c\balpha+\beta mc^2+e\phi(t,{\bi r}),
\]
where $\balpha=\gamma^0\bgamma$ and $\beta=\gamma^0$, are Dirac's hermitian matrices. In the original papers
Dirac uses a different notation for the above matrices but we have kept the today's more accepted one.
The potentials
are defined at point ${\bi r}$ and Dirac spinor $\psi(t,{\bi r})$ is a complex four-component
object also defined at point ${\bi r}$.

When computing the velocity of point ${\bi r}$ he obtains
\[
{\bi u}=\frac{d{\bi r}}{dt}=\frac{i}{\hbar}[H,{\bi r}]=c\balpha,
\]
irrespective of whether the particle is free or not. It writes on page 262 of his book \cite{Diracbook}: 
"\ldots{\it a measurement of a component
of the velocity of a free electron is certain to lead to the result $\pm c$. This conclusion is easily seen 
to hold when there is a field present}".

The total angular momentum of the electron with respect to the origin of the observer frame is
\[
{\bi J}={\bi r}\times{\bi p}+{\bi S},\quad {\bi S}=\frac{\hbar}{2}\pmatrix{\bsigma&0\cr 0&\bsigma},
\]
where ${\bi S}$, written in terms of Pauli matrices, represents the angular momentum with 
respect to the point ${\bi r}$. Both parts ${\bi r}\times{\bi p}$ and ${\bi S}$ are not conserved for 
the free electron. In the introduction of \cite{Dirac1}, Dirac writes: 
"{\it The most important failure of the model
seems to be that the magnitude of the resultant orbital angular momentum of an electron moving in an 
orbit in a central field of force is not a constant, as the model leads one to expect.}"

The spin part ${\bi S}$ satisfies in general
\[
\frac{d{\bi S}}{dt}=\frac{i}{\hbar}[H,{\bi S}]={\bi p}\times c\balpha={\bi p}\times{\bi u}
\]
even under the external interaction. 

The linear momentum is not along this velocity but is related to some average value: "{\it \ldots
the $x_1$ component of the velocity $c\alpha_1$, consists of two parts, a constant part $c^2p_1H^{-1}$,
connected with the momentum by the classical relativistic formula, and an oscillatory part, 
whose frequency is at least $2mc^2/h$,\ldots}". Point ${\bi r}$ is not the position of the CM. The frequency
predicted by Dirac for the motion of the point ${\bi r}$ is just twice de Broglie's postulated frequency.

When expanding the Hamiltonian in powers of ${\bi p}$, i.e., which could be intrepreted as the expression
of the energy in terms of the CM motion, he finds, in addition to the interacting term $e\phi$, two new 
interaction terms:
\[
\frac{e\hbar}{2mc}\pmatrix{\bsigma&0\cr 0&\bsigma}\cdot{\bi B}+\frac{ie\hbar}{2mc}\balpha\cdot{\bi E}.
\] 
He says: "{\it The electron will therefore behave as though it has a magnetic moment $({e\hbar}/{2mc}){\bf\Sigma}$,
and an electric moment $({ie\hbar}/{2mc})\balpha$. The magnetic moment is the just assumed in the spinning electron model.
The electric moment, being a pure imaginary, we should not expect to appear in the model. It is doubtful whether
the electric moment has any physical meaning
\ldots}".

In his book \cite{Diracbook} gives the same expression but he never mentioned, even in subsequent works like in the Nobel
conference \cite{DiracNobel},
the existence of this electric dipole. He analyzes the magnetic interaction but devotes no single
line to the electric dipole interaction, which has appeared on the same footing as the magnetic one. 
He disliked that the electron would have some property which
would imply a loss of spherical symmetry. The absolute value of this term is the charge $e$ times a distance
$\hbar/2mc$. This operator with the charge $e$ supressed, is the relative position operator of the CC 
with respect to the CM. It simply means that the above dipoles represent the electromagnetic 
electron structure with respect to the CM, a magnetic dipole and also an electric dipole. 
Dirac's spin operator is not a conserved
magnitude for a free particle and satisfies the dynamical equation of the spin with respect to the point where the
external force is defined, the CC. 

This point is moving at the speed of light. In Dirac's Nobel lecture  \cite{DiracNobel}
he says: "{\it It is found that an electron which seems to us to be moving slowly, must actually have a very 
high frequency oscillatory motion of small amplitude superposed on the regular motion which appears to us. As a result
of this oscillatory motion, the velocity of the electron at any time equals the velocity of light. \ldots one must believe
in this consequence of the theory, since other consequences of the theory \ldots, are confirmed by experiment.}"

The electron has two different centers separated by half Compton's wavelength. 
The rotation of the CC around the CM,
in the center of mass frame is at a frequency twice the frequency postulated by de Broglie.
It is a clear conclusion if Dirac would have read Marianne's thesis.

%%%%%%%%%%%%%%%%%%%%%%%%%%%%%%%%%%%
\section{The kinematical theory of spinning particles}
\label{kinematic}
%%%%%%%%%%%%%%%%%%%%%%%%%%%%%%%%%%%

Another conclusion of Marianne's thesis is that if this motion of the CC around the CM at the speed of light,
represents the natural motion of a free elementary particle, when we take a particle from a free initial state
to a free final state, only two things can happen. One possibility is that the intermediate states 
would be some kind of excited states such that part of the energy transfered to the particle during the interaction
is used to modify its internal structure, which is finally liberated when freed.
The other possibility is that the internal structure of the particle is not modified by the interaction
and all energy is used in the displacement of the particle and not in modifying its rotational
motion. If the object is not an elementary particle, we
know that the first possibility is a normal one. Atoms, molecules have excited states. Elementary
matter seems to have no excited states. Its rotational energy cannot be modified, contrary to a rigid body, 
so that changing its rotational state is a kind of producing an excited state, 
and these states are excluded for elementary particles.

We raise this idea to a fundamental principle which we propose to call the Atomic Principle \cite{Atomic}.
The difference between an elementary particle and bound systems of elementary particles is that
an elementary particle has no excited states. From the classical point of view, in a variational approach 
of systems of a finite number of degrees of freedom, we have no way to describe how a particle
can be transformed into several particles. A classical particle can never be divided and it can 
never be deformed.
At every time of its evolution remains the same and we can always find some inertial observer who describes
the particle in the same state as the initial state. If the particle changes its rotational
motion this cannot be compensated by a transformation to a new inertial refence frame. 
In a classical variational description, the manifold of the
boundary states of an elementary particle is necessarily a homogeneous space of the kinematical 
group associated to the restricted
relativity principle. In the quantum case, the Hilbert space of its free vector states is, according to Wigner, 
the representation space of an {\it irreducible representation} of the kinematical group \cite{Wigner}.
This definition has been used to analyze different models of elementary spinning and spinless particles
\cite{Rivasbook}. 

The initial state of the point particle, from the variational point of view, is described
by $x_1\equiv\{t_1,{\bi r}_1\}$ and the final state by $x_2\equiv\{t_2,{\bi r}_2\}$, so that this manifold $X$, 
the spacetime, is clearly a homogeneous space of either the Galilei or Poincar\'e group. Then a point particle 
is an elementary particle according to this definition. 
But it is a spinless particle. 

If we postulate a restricted relativity principle by assuming that the 
kinematical group of spacetime symmetries is only the group of spacetime translations, i.e., that physical 
laws are invariant only under translations, this manifold will be the largest 
homogeneous space of this group, and the most structured elementary particle we can define under this assumption.
The free Lagrangian of an elementary particle will be a function of $L_0(t,{\bi r},{\bi u})$, and invariance
under translations means that the most general free Lagrangian will be an arbitrary function of the velocity
of the point ${\bi u}$.

If we now enlarge the kinematical group to include also the spatial rotations, we also have orientation variables 
to describe the boundary states of our elementary particle, then our most structured 
particle will have as boundary states the manifold spanned by $x\equiv\{t,{\bi r},\balpha\}$, i.e., a point ${\bi r}$ and a cartesian frame
linked to point ${\bi r}$, with an orientation described by the variables $\balpha$ of a suitable parameterization
of the rotation group. The most general free Lagrangian will be a function of $L_0(t,{\bi r},\balpha,{\bi u},\bomega)$
where $\bomega$ represents the angular velocity of the cartesian frame. Invariance under translations and rotations imply
that it is only a function of the velocity of the point ${\bi u}$ and of angular velocity in the form of an
arbitrary function $L_0(u^2,\omega^2,{\bi u}\cdot{\bomega})$. The point ${\bi r}$ will represent the location
of the center of mass and also the point where the external interactions are defined. Dynamical equations will  
be second order differential equations. The elementary particle is described as a kind of 
a spherically symmetric rigid body, because dependence on the angular velocity is in 
the form of a function of $\omega^2$. We cannot make any assumption about its shape or size,
but only that the object has a unique moment of inertia $I$, and therefore the description
of a gyration radius. This elementary
particle has spin, ${\bi S}=I\bomega$, i.e., 
angular momentum with respect to the center of mass.

If we also assume that physical laws are also invariant under a change of reference frame to another arbitrary observer
moving with constant velocity, we can get different kinematical groups \cite{BacryLL}. Among them,
the Galilei and Poincar\'e groups.

The Atomic Principle states that we can enlarge the manifold of boundary states 
to the largest homogeneous space of the kinematical group, but no more than that. 
In the Galilei case the largest one is the group
itself. The group parameters become the classical variables which define the possible boundary states
of the elementary particle. These are $x\equiv(t,{\bi r},{\bi u},\balpha)$, 
the time $t$, the position of a point ${\bi r}$, 
the velocity of this point ${\bi u}$, and the orientation $\balpha$ of a Cartessian frame located
at ${\bi r}$. An elementary particle is a localized system at a point ${\bi r}$ and also orientable whose
spatial orientation is described by the orientation of its local frame ${\balpha}$. 
Elementary matter moves and rotates. 
The Lagrangian will be a function of these variables and of the next order time derivative of them, i.e.,
it will depend on the acceleration of the point ${\bi r}$ and on the angular velocity of the Cartesian frame.
The dynamical equations of the point ${\bi r}$ are fourth order differential equations. This dependence
of the Lagrangian on the acceleration prevents the linear momentum to be collinear with the velocity
of the point ${\bi r}$, but it is in ${\bi r}$ where the external potentials are defined. The particle
has spin which comes from two different terms: the dependence of the Lagrangian on the acceleration and
the dependence on the angular velocity. We describe
an spinning object which has a CC different than the CM. In this case, the point ${\bi r}$ is not the center
of mass of the system but it is the point where the external potentials are defined. But this nonrelativistic
description does not supply models with unreachable velocities, like the one suggested in Marianne's thesis.

In the Poincar\'e case, the largest manifold is spanned by the same variables but now with three possible restrictions,
$u<c$, $u=c$ or $u>c$. By Marianne's thesis, it is the manifold with $u=c$ which seems to describe her model.
It is the only model which, when quantized, satisfies Dirac's equation \cite{RivasDirac}. 
We can say that Dirac's equation
is satisfied by classical material systems whose internal structure cannot be modified.
These particles have a unique internal rotation and therefore, a unique angular momentum.
In \cite{Rivasbook}
we have shown that the quantum expression of the observable $e({\bi r}-{\bi q})$ 
is exactly Dirac's electric dipole operator.

In general, the free Lagrangian for this system will be a function $L_0(t,{\bi r},{\bi u},{\bi a},\balpha,\bomega)$
of time, position of a point ${\bi r}$, its velocity ${\bi u}$ and acceleration ${\bi a}$, 
and of its spatial orientation $\balpha$ and its time derivative, the
angular velocity $\bomega$. Invariance under translations and rotations imply that finally is a function
of only $L_0({\bi u},{\bi a},\bomega)$. For Lagrangians depending on the acceleration, 
the linear momentum takes the form
\[
{\bi p}=\frac{\partial L_0}{\partial{\bi u}}-\frac{d}{dt}\left(\frac{\partial L_0}{\partial{\bi a}}\right).
\]
If we call ${\bi U}={\partial L_0}/{\partial {\bi a}}$, and
 ${\bi W}={\partial L_0}/{\partial {\bomega}}$, the angular momentum with respect to the origin of observer's frame becomes:
\[
{\bi J}={\bi r}\times{\bi p}+{\bi u}\times{\bi U}+{\bi W}={\bi r}\times{\bi p}+{\bi S},
\]
so that the observable ${\bi S}$ is the angular momentum with respect to the point ${\bi r}$. 
It contains two parts. One ${\bi Z}={\bi u}\times{\bi U}$, coming from the dependence of 
the Lagrangian on the acceleration, which is orthogonal to the velocity ${\bi u}$ and also to the vector ${\bi U}$,
which in the CM frame ${\bi U}\sim -{\bi a}$ has a direction opposite to the acceleration. Its relative 
orientation is depicted
in figure \ref{electronCM}. The second part ${\bi W}$, which is pointing along $\bomega$, related to the rotation
of the particle local frame. But point ${\bi r}$ is not the CM of the particle because ${\bi p}$ is not lying
along the velocity ${\bi u}$. Therefore, this angular momentum satisfies the dynamical 
equation $d{\bi S}/dt={\bi p}\times{\bi u}$, even under interaction,
like Dirac's spin operator. The point ${\bi r}$ is the point where the external potentials are defined. The dynamics
describes the evolution of the CC. 

Dirac's Hamiltonian, from the classical point of view, takes the form
\[
H={\bi p}\cdot{\bi u}+\frac{1}{c^2}{\bi S}\cdot\left(\frac{d{\bi u}}{dt}\times{\bi u}\right).
\]
It is the decomposition of the mechanical 
energy of the electron in two parts, one the translational energy related to the mechanical linear momentum,
${\bi p}\cdot{\bi u}$, which can be zero in some frames, and the rotational internal energy related to the spin with respect to the CC.
This second part never vanishes, because the spin ${\bi S}$ is along the 
direction of the cross product of the
acceleration and velocity. It is positive for the particle and negative for the antiparticle.
When quantizing the system the first part is transformed into ${\bi p}\cdot c\balpha$ 
and the second becomes the $\beta mc^2$ part for the free particle.

\cfigl{electronCM}{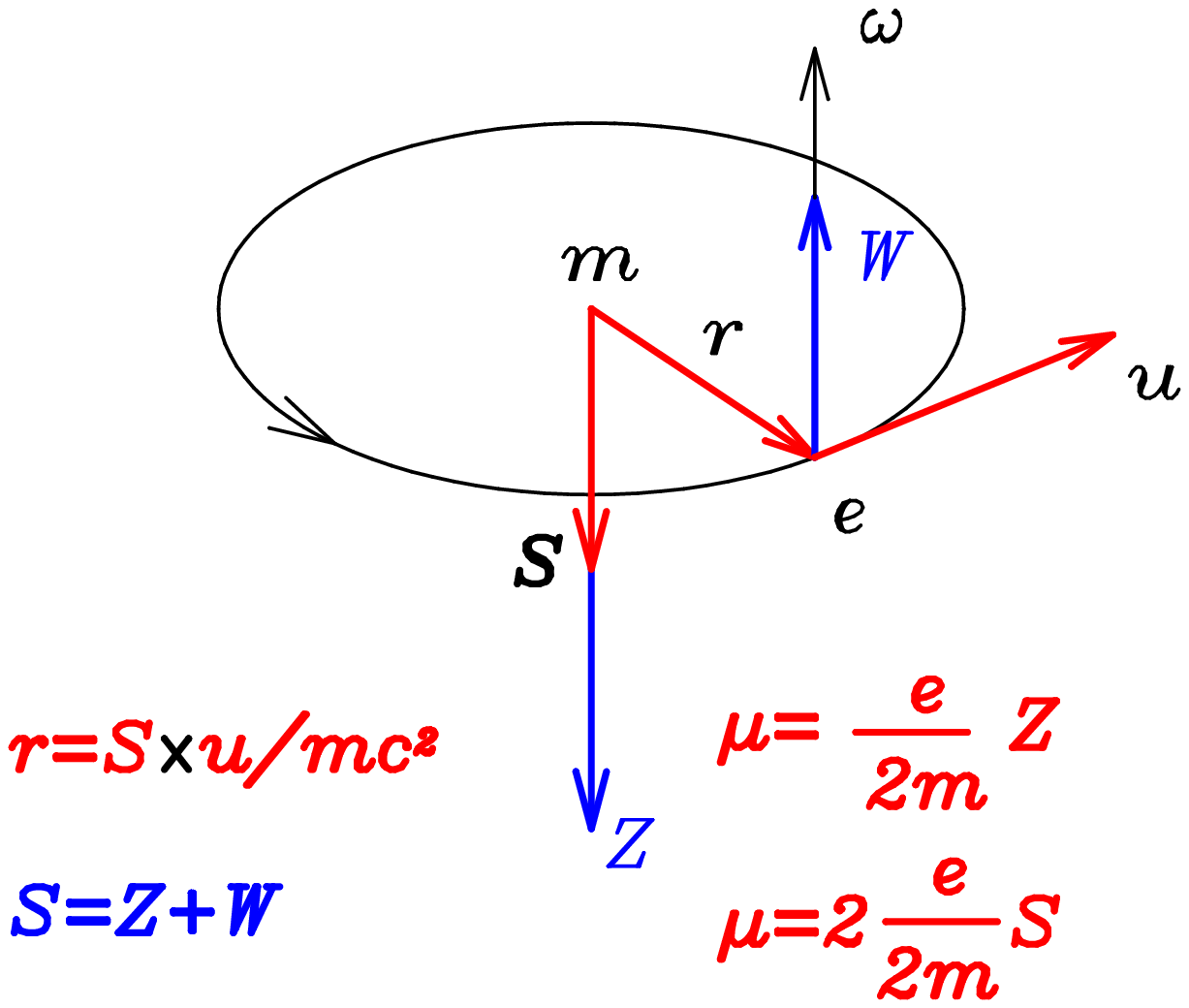}{Motion of the CC,  ${\bi r}$, where the charge $e$ is located, 
around the CM at the origin in this frame ${\bi q}=0$.
The particle also has located at CC a rotating frame which rotates with angular velocity $\bomega$. This local
frame, for instance its Frenet-Serret triad, has not been depicted. The angular momentum with respect to any point in this frame is the same ${\bi S}$, 
because ${\bi p}=0$. It consists of two terms ${\bi S}={\bi Z}+{\bi W}$. One, ${\bi Z}$ coming from the dependence
of the Lagrangian on the acceleration and which, for the particle $(H>0)$, 
has the relative orientation depicted in the figure. For the antiparticle, $(H<0)$, has the opposite orientation.
When quantizing the model it quantizes with integer values. For an elementary particle it takes the minimum nonvanishing value
$Z=\hbar$. Another ${\bi W}$, in the same direction as the angular velocity of the rotating local frame, 
related to the dependence of the Lagrangian on the angular velocity. These two parts have opposite orientations, 
so that the total spin ${\bi S}$ has a value smaller than the value of ${\bi Z}$. 
The ${\bi W}$ quantizes with the minimum nonvanishing value, $W=\hbar/2$. When quantized this model, $S=\hbar/2$. 
The magnetic moment $\bmu$ is produced by the orbital motion of the charge and thus related 
to the ${\bi Z}$ part of the spin by the usual relation.
When related to the total spin ${\bi S}$ predicts a gyromagnetic ratio $g=2$. 
This object has also
an oscillating electric dipole ${\bi d}=e{\bi r}$, with respect to the CM.}

The description of an elementary particle as only a strict mathematical point, 
is a very restricted formalism. It is equivalent to consider that the laws of physics are only invariant 
under spacetime translations. Special relativity assumes Poincar\'e invariance of physical laws, 
and therefore the most structured object which fullfils with the requirement that its possible states
are only kinematical modifications of any one of them, i.e., that
its internal structure is not modified by any interaction, implies the above description, 
where the point ${\bi r}$,
which represents the point where the external interaction is defined, is a different point than the CM.
A relativistic charged spinning particle has a CC and a CM which are different points.
 
%%%%%%%%%%%%%%%%%%%%%%%%%%%%%%%%%%%
\section{Some predicted effects}
%%%%%%%%%%%%%%%%%%%%%%%%%%%%%%%%%%

By using this relativistic model where the CC moves at the speed of light we have shown different
physical effects which we summarize in what follows. We shall outline the main features leaving
for the interested readers, to go through the original publications.

The description of the motion of the CC at the speed of light, in the center of mass frame, 
is depicted in the figure \ref{electronCM}.

%%%%%%%%%%%%%%%%%%%%%%%%%%%%%%%%%%%
\subsection{The gyromagnetic ratio of the electron}

The double structure of the spin ${\bi S}={\bi Z}+{\bi W}$, with ${\bi S}$ pointing in the same direction
as the part ${\bi Z}$ but opposite to the direction of ${\bi W}$, means that the absolute value of $S<Z$. 
But the ${\bi Z}$ part is coming from
to the orbital motion of the charge and therefore is related to the magnetic moment of the particle. 
From the classical point of view
the values of $S$, $Z$ and $W$ are unrestricted. But when we quantize the model, to give stability to the particle and no radiation
to the motion of the charge and assume that the corresponding
values of $Z$ and $W$ are the smallest nonvanishing values compatible with their quantum mechanical structure, we get
$Z=\hbar$ and $W=\hbar/2$, and thus $S=\hbar/2$ pointing along ${\bi Z}$ \cite{Rivasetal}.
The magnetic moment in the CM frame takes the form
\[
\bmu=\frac{e}{2m}{\bi Z}=2\frac{e}{2m}{\bi S},\quad g=2,
\]
when expressed in terms of the total angular momentum in this frame.

The total $S=\hbar/2$ can also be obtained from a different analysis. 
The radius of this motion is $R=S/mc$, and the angular velocity is $\omega=mc^2/S$. 
When analyzed this model in the CM frame, instead of describing
a system of six degrees of freedom, we are left with only three. These are the coordinates $x$ and $y$ of 
the CC in the plane orthogonal to the spin if this is taken along $OZ$ axis, and also the phase $\alpha$ of the rotation
of its Frenet-Serret triad. But this $\alpha$ is the same as the phase of the orbital motion. 
Finally, because the motion is 
a circle at a constant speed, only one of the variables, let us say $x$, is the only 
independent degree of freedom in this frame. From a strict mechanical point of view and in the CM frame
the system is a one-dimensional harmonic oscillator of frequency $\omega$ in its ground state. 
Since it is an elementary
particle it has no excited states so that its total energy in the CM frame $mc^2$ is the ground state energy
of the quantized harmonic oscillator $\hbar\omega/2$, so that $S=\hbar/2$, when quantized.

%%%%%%%%%%%%%%%%%%%%%%%%%%%%%%%%%%%
\subsection{Are the spin and magnetic moment of the electron, parallel or antiparallel vectors?}

In figure \ref{electronCM} we have depicted the motion of the CC of a system for which the Hamiltonian $H>0$,
i.e., what is known as the particle. For the antiparticle $(H<0)$ we have the time reversed motion while maintaining
the same total spin. But the formalism does not fix the sign of the charge. For the particle it can be either $e>0$
or $e<0$. If we consider that the particle is the electron and thus $e<0$, the magnetic
moment of the electron ${\bmu}$, corresponding to this motion, is pointing down in the same direction as the total spin ${\bi S}$. For the antiparticle,
the motion is reversed and also the sign of the charge, so that we get that for the positron spin and magnetic moment
are also parallel. If we had chosen the opposite sign for the charge of the particle, spin and magnetic moment will be antiparallel,
but also for the antiparticle. We arrive at the conclusion that Dirac particles and antiparticles must have the same
relative orientation between the spin and magnetic moment. 

This conclusion seems to hold in the case of the positronium ground state. It is a system of a positron and an electron
in a $L=0$ orbital angular momentum state. The magnetic moment and the spin of the positronium 
ground state are both zero thus confirming that in this state particle and antiparticle have their spins pointing
in opposite directions and also for their magnetic moments. But we do not know if for each object 
they are parallel or antiparallel.

The same thing happens for the neutral pion $\pi^0$. It is considered as a vector state $\pi^0=(u\bar{u}+d\bar{d})/\sqrt{2}$
of the two pairs of up and down quarks and antiquarks $u\bar{u}$ and $d\bar{d}$, respectively. 
These quarks have different masses and charges, so that they have different magnetic moments. 
Nevertheless, the spin and magnetic moment
of $\pi^0$ are both zero, so that each pair $u\bar{u}$ and $d\bar{d}$ must have zero spin and 
zero magnetic moment.
Quarks and their antiquarks must have the same relative orientation between the spin and 
the magnetic moment \cite{magnetic}.

The sign of the charge as well as the concept of particle and antiparticle is a convention. But the formalism
does not give a relationship between them. Only that the charges are opposite to each other as well as 
the value of the observable $H$. 
Once the sign of the charge is fixed, the magnetic moment is determined if the motion of the charge is known.
Therefore we have to determine the relative orientation between spin and magnetic moment, experimentally.
We have proposed two experiments to determine for electrons this relationship \cite{magnetic}. 

%%%%%%%%%%%%%%%%%%%%%%%%%%%%%%%%%%%
\subsection{The electron clock}

If the electron corresponds to this model it has an internal frequency and therefore can be interpreted 
as a clock. Is this frequency the one postulated by de Broglie $\omega=mc^2/\hbar$ or the one obtained by Dirac, twice de Broglie's frequency?
We have proposed to enlarge the energy range \cite{clock}, 
of a previous experiment performed by Gouan\'ere et al. \cite{Gouanere}
to determine the peaks of a resonant scattering of a beam of electrons accelerated to some accurate velocity, and interacting
with the atoms of a sylicon crystal.

The idea is that this periodic in time internal motion, when the particle is displaced, also corresponds to a spatial
periodic motion of a coherent beam of particles. 
We can talk about the "{\it wavelength}" of the particle as the distance covered by the CM 
during one turn of the CC.
If this distance is commensurable with the separation of the atoms of a crystal, there will be a resonant 
transfer of transversal linear momentum to the electrons in the beam, 
so that the intensity of the outgoing beam of electrons in 
the forward direction will decrease, for some specific incoming velocities.

Measuring the velocity of the electron beam for the resonant peaks, 
will give us information about the internal frequency $\omega$.

%%%%%%%%%%%%%%%%%%%%%%%%%%%%%%%%%%%
\subsection{Spin polarized tunneling} 

If the electron has two separate centers, and tries to penetrate a potential barrier, a force will be 
exerted whenever the CC is under the external field. We have analyzed the interaction of such a particle with a
triangular barrier \cite{tunneling}, i.e., a barrier which has a region where the force is opposite to the motion
of the particle and brakes its motion, and another, where the force points in the forward direction. 
If the electron is polarized with the spin along the direction of motion,
when the CC penetrates into the barrier, the same happens to the CM, and conversely. In this case no tunneling
appears. But if the electron is polarized in a direction orthogonal to the motion, the field produces a work when
the CC is inside the barrier and therefore the kinetic energy decreases. 
But when the motion of the CC is outside the barrier, although the CM could be inside,
the kinetic energy remains constant and thus the CM penetrates more into the barrier. 
It may happen that this penetration is sufficient for the CC to reach the other side of 
the barrier, where the external force has the direction of the motion. 
In that case, while the CC is in the extracting
region of the potential, the kinetic energy of the particle increases, and thus the CM penetrates more
into the barrier. 
What we have shown is that for every kind of a barrier, there exists an energy, below the top of the potential,
for which the classical particles cross the barrier.

For this effect to happen, one of the two regions of the triangular potential has to have a depth of the order
of the separation between CC and CM. Another condition is that the electron has to be polarized transversaly
to the motion. This is known in the literature as {\it spin polarized tunneling}. 

Magnetoresistive materials are materials of a very high resistivity. They are syntherised materials formed
by small grains of a substance, hold together by compression. 
The current is produced by the jump, from grain to grain,
of the conducting electrons trying to overcome a kind of triangular potential between grains. It is said
that conductivity through the grains is produced by tunneling, 
thus justifying the high resistivity of the sample. 
But when we introduce a magnetic field in a direction orthogonal to the current, 
the resistivity almost decreases to zero. Electrons become polarized in a direction orthogonal 
to the current and the spin polarized effect happens. 
All electrons with an energy below the top of the potential, but above the minimum energy for crossing,
go through the barrier like in a normal conductor.
If the electrons are polarized at random, only a few percentage crosses the barrier, and therefore
the conductivity is low, but when all are polarized, the effect is enhanced and the sample becomes a good conductor.

%%%%%%%%%%%%%%%%%%%%%%%%%%%%%%%%%%%
\subsection{The formation of bound pairs of electrons}

If the electron has two separate centers and we consider the interaction of two electrons with their spins
parallel we can obtain a bound state of them. This can be obtained
provided some separation $a$ below Compton's wavelength between the CM's, a 
relative velocity of the CM's below $0.01c$ and opposite internal phases of the CC motion \cite{dynamical}.

In the figure \ref{bound} we see how a repulsive force between the charges implies an attractive force for the motion
of the corresponding CM's.
%\vspace{-1.5cm}

\cfigl{bound}{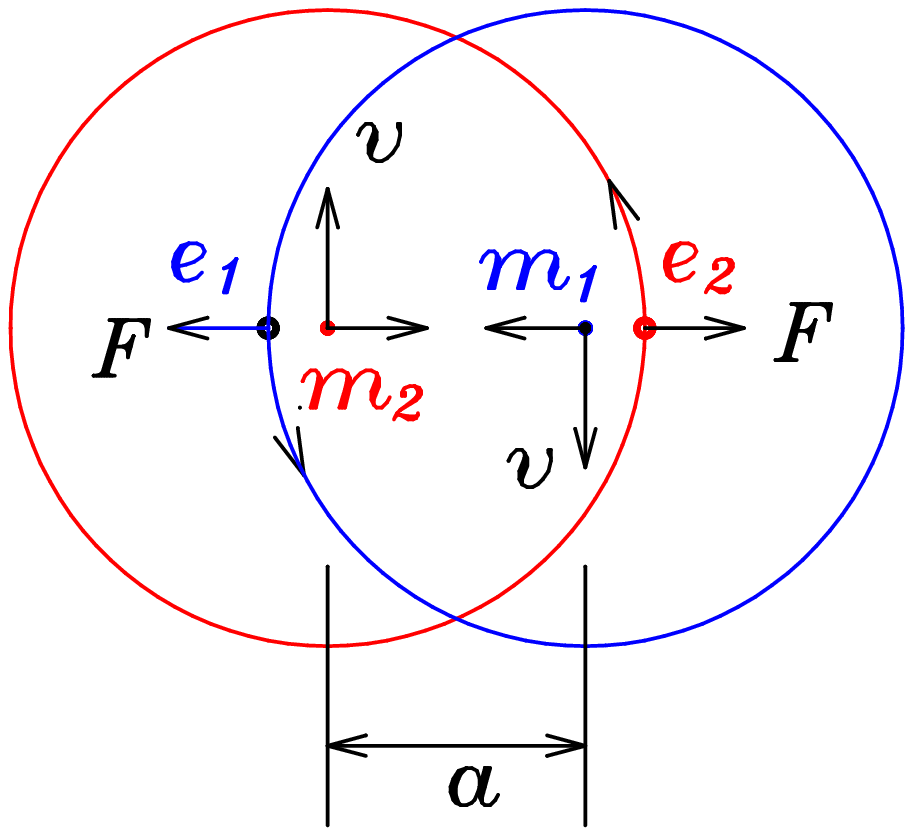}{Initial position and velocity of the CM's and CC's for a bound motion 
of a two-electron system with parallel spins. The circles would correspond to the tentative
trajectories of the charges if considered free. The interacting force $F$ is computed
in terms of the separation between the charges and their velocities. These forces defined at each charge position,
are also depicted in the corresponding CM to analyze its dynamics. A repulsive force between charges
is an attractive force between the CM's provided some kinematical conditions are fulfilled.}

Classical physics does not forbid that two electrons with the spins parallel form a metastable bound state,
which, when quantized, corresponds to a boson of spin 1. The two electrons are in the same kinematical state, and only
their phases have to be opposite to each other for the state to be stable. Is this a classical way to 
circumvent the Pauli quantum exclusion principle?

%%%%%%%%%%%%%%%%%%%%%%%%%%%%%%%%%%%
\subsection{The proton spin crisis}

If quarks are Dirac particles, the above analysis of the electron can be applied to a quark where now ${\bi r}$
represents the location of the interacting charge of the quark, the colour charge. 
It can be either the electric or the colour charge, but in any case a different point than the CM.

We see in figure \ref{proton} that if the spin of the proton is the angular momentum of the three quarks with respect to the
center of mass of the system, then the addition of the three Dirac operators, which correspond to the addition of the three
angular momenta with respect to the corresponding CC's, would never produce the total angular momentum of the proton.
If, as assumed, the three quarks in the ground state of the proton, are in a $L=0$ orbital angular momentum, what
we have to add are the three $S_{CM}$ for each quark. Therefore, a term of the form $({\bi r}-{\bi q})\times{\bi p}$,
for each quark is lacking in the computation of the proton spin. But for the position vector operator
${\bi r}-{\bi q}$ we have to use Dirac's electric dipole operator, with the charge $e$ deleted \cite{lacking}.

\cfigl{proton}{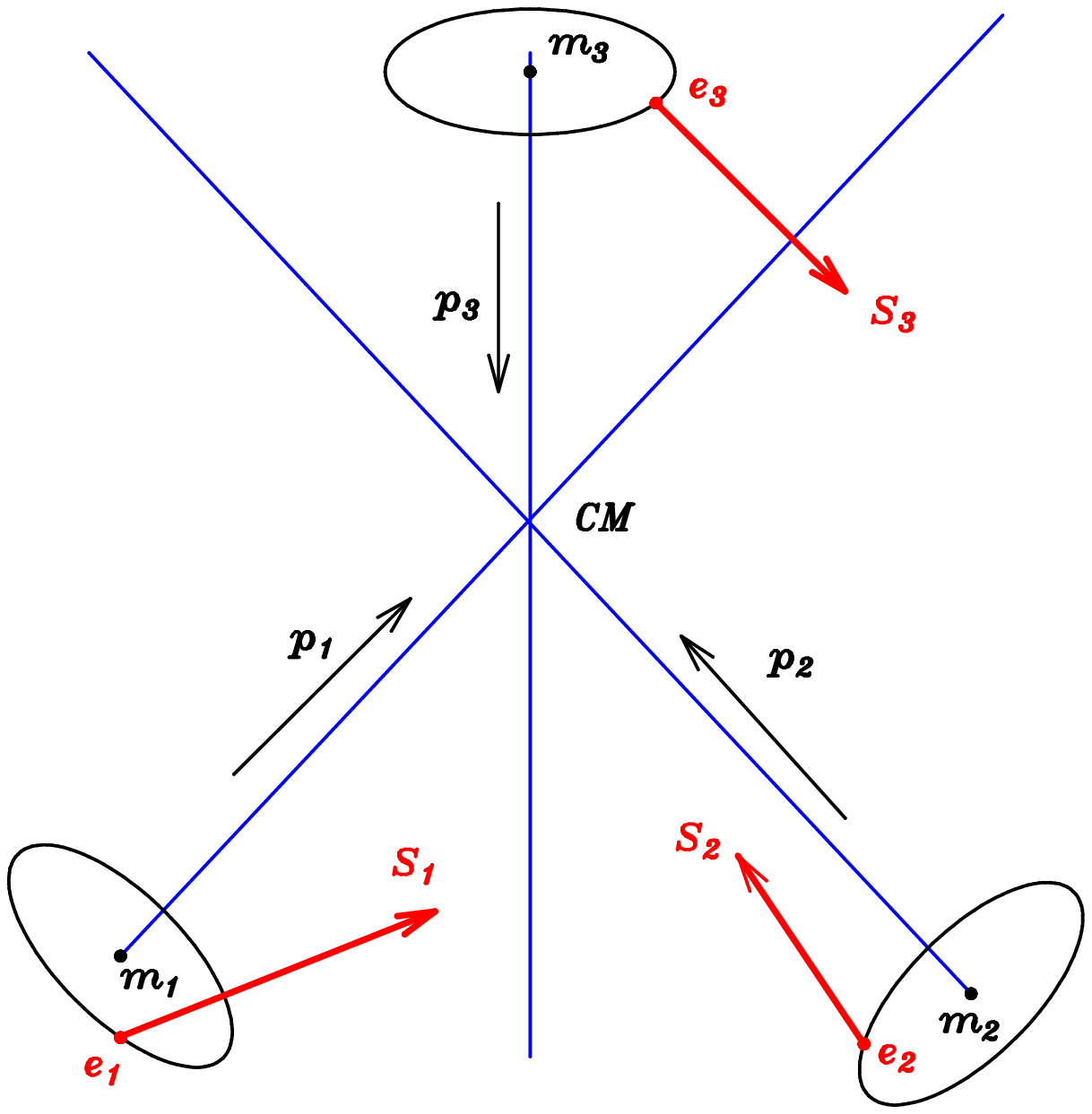}{Proton structure as a bound system of three Dirac particles in a $L=0$ state. The motion
of the 3 CM's lie on a plane. The addition of the three angular momenta w.r.t. the CC's, ${\bi S}_1$, ${\bi S}_2$ and ${\bi S}_3$,
can never give rise to the angular
momentum of the system w.r.t. the common CM.}

%%%%%%%%%%%%%%%%%%%%%%%%%%
\ack{I would like to thank J.M. Aguirregabiria for the use of his Dynamics Solver program \cite{DSolver}, 
which has been used in many of the mentioned physical analysis and to 
Juan Barandiaran for his skill in producing a Mathematica
package \cite{Math} with which we have analyzed the classical interaction of two spinning electrons, its scattering
and the formation of bound pairs.}

%%%%%%%%%%%%%%%%%%%%%%%%%%

\newpage

\end{document}